\documentclass{josis}
\usepackage{hyperref}
\usepackage{nameref}
\usepackage[hyphenbreaks]{breakurl}
\usepackage{booktabs}
\usepackage{stmaryrd}
\usepackage[T1]{fontenc}
\usepackage{cite}

\usepackage{algorithm}
\usepackage{algpseudocode}

\usepackage{graphicx}
\usepackage{subcaption}
\usepackage{float}

\usepackage[table]{xcolor}
\usepackage{amssymb,amsmath}

\josisdetails{%
   number=N, year=YYYY, firstpage=xx, lastpage=yy,
  doi={10.5311/JOSIS.YYYY.II.NNN},
   received={December 24, 2015},
   returned={February 25, 2016},
   revised={July 13, 2016},
   accepted={September 5, 2016}, }

\makeatletter
\let\UrlSpecialsOld\UrlSpecials
\def\UrlSpecials{\UrlSpecialsOld\do\/{\Url@slash}\do\_{\Url@underscore}}%
\def\Url@slash{\@ifnextchar/{\kern-.11em\mathchar47\kern-.2em}%
    {\kern-.0em\mathchar47\kern-.08em\penalty\UrlBigBreakPenalty}}
\def\Url@underscore{\nfss@text{\leavevmode \kern.06em\vbox{\hrule\@width.3em}}}
\makeatother

\hypersetup{
colorlinks=true,
linkcolor=black,
citecolor=black,
urlcolor=black
}

\runningauthor{\begin{minipage}{.9\textwidth}\centering Fleischmann, Vybornova\end{minipage}}
\runningtitle{A shape-based heuristic for the detection of urban block artifacts}

\begin{document}

\title{A shape-based heuristic for the detection of urban block artifacts in street networks}

\author{Martin Fleischmann}\affil{Department of Social Geography and Regional Development, Charles University, Czechia}
\author{Anastassia Vybornova}\affil{NEtworks, Data and Society (NERDS), Computer Science Department, IT University of Copenhagen, Denmark}

\maketitle


\begin{abstract} Street networks are ubiquitous components of cities, guiding their development and enabling movement from place to place; street networks are also the critical components of many urban analytical methods. However, their graph representation is often designed primarily for transportation purposes. This representation is less suitable for other use cases where transportation networks need to be simplified as a mandatory pre-processing step, e.g., in the case of morphological analysis, visual navigation, or drone flight routing. While the urgent demand for automated pre-processing methods comes from various fields, it is still an unsolved challenge. In this article, we tackle this challenge by proposing a cheap computational heuristic for the identification of ``face artifacts'', i.e., geometries that are enclosed by transportation edges but do not represent urban blocks. The heuristic is based on combining the frequency distributions of shape compactness metrics and area measurements of street network face polygons. We test our method on 131 globally sampled large cities and show that it successfully identifies face artifacts in 89\% of analyzed cities. Our heuristic of detecting artifacts caused by data being collected for another purpose is the first step towards an automated street network simplification workflow. Moreover, the proposed face artifact index uncovers differences in structural rules guiding the development of cities in different world regions. 

\end{abstract}

\section{Introduction}
\label{sec:intro}
Cities have been the object of human inquiry for thousands of years \cite{vitruvius_vitruvius_1999} and for a wide range of scientific fields, such as growth dynamics, population development, and spatial structure. Within the last 50 years, powered by the emergence of Big Data and computational power, data-driven approaches to the study of cities have gained unprecedented importance \cite{arcaute_recent_2022}. For studies concerned with aspects of urban form, such as city structure \cite{el_gouj_urban_2022}, place connectivity \cite{domingo_graph-based_2019}, accessibility \cite{rhoads_inclusive_2023}, resilience of the built environment \cite{sharifi_resilient_2019}, and liveability \cite{wang_re-examining_2022}, street network data has proven to be a particularly useful point of departure. Street networks are line-based abstractions of the street space, containing information on the connections between intersections and street segments. They are popular objects of urban analyses due to their explanatory power, direct relationship to an extensive field of graph theory, and simplicity (digitizing a street network is much easier than digitizing buildings). The feasibility of studies that take street networks as a point of departure has greatly increased with open source geospatial data becoming available on platforms like OpenStreetMap (OSM) \cite{openstreetmap_contributors_openstreetmap_2023}, allowing a wide range of applications \cite{arcaute_recent_2022}, from transportation network design \cite{farahani_review_2013}, assessment of urban sprawl \cite{barrington-leigh_global_2020} or evolution and distribution of street patterns \cite{boeing_multi-scale_2018, boeing_off_2021} to the classification of urban form \cite{araldi_street_2019,fleischmann_methodological_2021}.

However, street network datasets vary significantly in their level of detail and data quality \cite{haklay_how_2010, boeing_street_2022}. Whether a street network data set is fit for purpose depends to a large extent on the specific application. A frequent data processing challenge, common to many street network applications, is to reduce the level of detail from one use case (e.g.~traffic routing) to the other (e.g.~morphological analysis) without losing relevant information or introducing new imprecisions. For example, traffic routing applications require adequately represented directionality of edges (street segments) \cite{luxen_real-time_2011}, while studies on urban morphology \cite{venerandi_form_2017, dibble_origin_2019} or on urban air space \cite{badea_limitations_2021, vidosavljevic_metropolis_2021} aim to reflect the space in between buildings and its perceptional configuration and thus do not require the edges to be directed, and do not want transportation geometries like roundabouts included. In such cases, manually deciding which information to keep and which to aggregate is prohibitively time-consuming and therefore not feasible at scale, while implementing an algorithm to automate this decision is a challenging task (see Figure~\ref{fig:faceartifacts}). 

In short, when it comes to automated pre-processing methods for street network data, demand from various fields is high, but supply is currently still low. Here, we take a first step towards a fully automated street network simplification method by developing a computationally cheap heuristic for the identification of ``face artifacts'', i.e., polygons that are enclosed by transportation network edges but do not represent urban blocks. As described in detail below, face artifacts are a hindrance both for studies that are concerned with the morphology of the street network itself and for studies that investigate the morphology of urban blocks, commonly conceptualized as the polygons \textit{enclosed by} the street network. We therefore pose the following research question:

\begin{center}
\textit{How can face artifacts in an urban street network be computationally identified?}
\end{center}

\noindent In this article, we propose a method to answer this question, and test the proposed method’s universality on case studies from across the globe. The rest of the paper is organized as follows: next, we provide a precise problem description, followed by a review of previous work and terminology. We then introduce our methodology, proposing a cheap computational heuristic that allows the automated identification of face artifacts based on the geometric shape of areas enclosed by street edges. We apply the heuristic to 131 cities across the globe, and present and evaluate the results. We conclude by discussing our contribution and outlining potential further steps towards a fully automated street network simplification method.

\subsection*{Problem description: Face artifacts}
\begin{figure}[t]
    \centering
    \includegraphics[width=\linewidth]{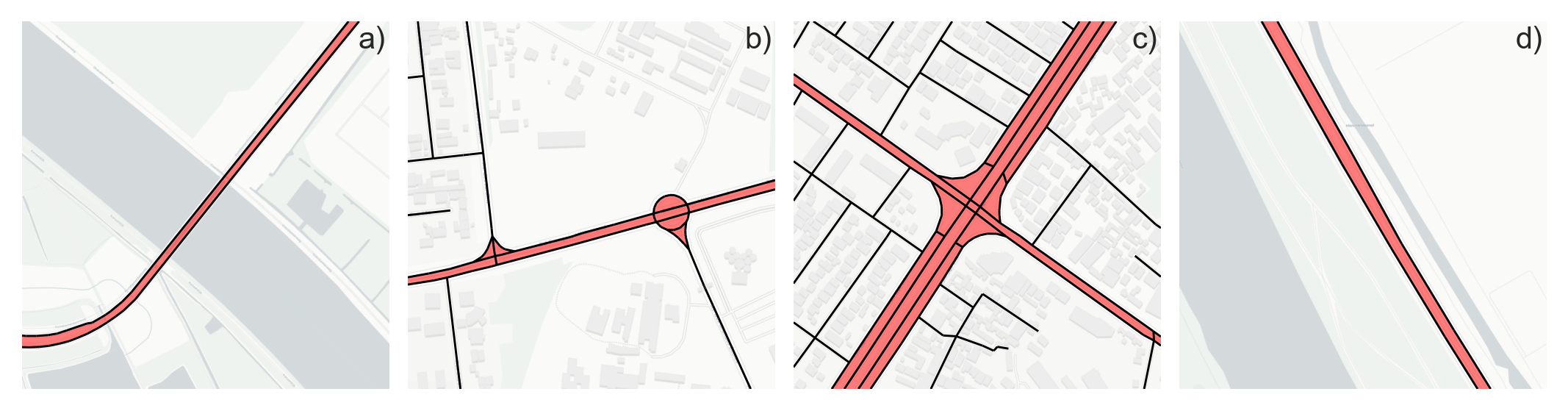}
    \caption{a) Bridge, Amsterdam; b) Roundabout, Abidjan; c) Intersection, Kabul; d) Motorway, Vienna. Polygons classified as face artifacts are shown in red, and the OSM street network (without service roads) is shown in black. Face artifacts are polygons enclosed by street network geometries (in the case of OSM, lane centerlines) that do not represent morphological urban blocks, but instead are a result of detailed transportation-focused mapping of the streetscape. Map data \textcopyright OpenStreetMap contributors \textcopyright CARTO}
    \label{fig:faceartifacts}
\end{figure}

Each geospatially encoded street network comes with a certain level of detail and a focus on specific elements of street space. While all attempt to capture primarily connectivity, which is the primary purpose of streets, the resulting geometries and their graph representations can vastly differ. On one hand, we have networks where every traffic lane and every intersection have been meticulously digitized with their respective attributes of directionality, hierarchy, and speed limits; this type of graph representation is needed for routing applications \cite{luxen_real-time_2011, wang_estimating_2011, schroder_graphhopper_2017}. On the other hand, we have networks that attempt to capture human perception of space and certain aspects of navigation in a city, where detailed information on intersection geometries usually makes little sense and is omitted \cite{araldi_street_2019, ballo_modeling_2023}. While each of these graph representations has its own use cases, the former is more frequently available; therefore, researchers often need to start from a detailed network and derive a simplified one before starting an analysis. The difference between a transportation-focused and a perception-focused network can be better understood by examining the areas enclosed by network edges, which is a commonly used proxy for urban blocks \cite{barthelemy_paths_2017, grippa_mapping_2018, fleischmann_measuring_2021, shpuza_shape_2022}. In graph theory, the polygons enclosed by the network edges in a planar space are called \textit{(graph) faces} \cite{brightwell_representations_1993}. A detailed look at these face polygons for a given street network often reveals artifacts of transport-focused geometry -- polygons that do not capture urban blocks. Figure~\ref{fig:faceartifacts} illustrates this issue in OpenStreetMap, which is a widely used open source for global street network data \cite{grinberger_osm_2022} with a transportation-focused mapping approach \cite{openstreetmap_contributors_highways_nodate}.

\noindent The face polygons colored in red are not urban blocks enclosed by streets; rather, they appear in the network due to the representation of dual carriageways as separate network edges (lane centerlines). This way of network representation is suitable for traffic routing but can pose problems for other applications. When the goal is to generate polygons that are representative of urban blocks and at the same time to ensure that the graph represents the morphological network rather than the transportation network, we suggest calling such face polygons ``face artifacts'', as they occur only as a result of the data preparation model not suited for this goal. 

Face artifacts pose a twofold problem. First, in studies concerned with urban form, they introduce a false signal into the distribution of urban shapes and distort the polygon representation of the actual urban blocks. Second, in studies concerned with the properties and patterns of the street network, face artifacts introduce superfluous network edges, thus distorting all network metrics based on node degree and/or shortest path computations. A further aggravating factor is that the extent to which face artifacts distort results depends on the analysis conducted and cannot be quantified without prior identification of such polygons or superfluous edges. Thus, no matter whether one is interested in the topology of the urban street network or the morphology of urban shapes enclosed by the network, face artifacts might need to be removed as part of data preprocessing and replaced by single network edges. Human manual identification of face artifacts would be unambiguous but prohibitively costly, not scalable, and not entirely reproducible. Although many authors have already pointed out this issue in a wide range of contexts (see section below), a fully automatized approach to identifying and potentially removing face artifacts is, to our knowledge, still non-existent.

\subsection*{Previous work}

While face artifacts are a commonly known problem in the research community, there is a lack of coherent terminology for the phenomenon. Previous studies have referred to the same issue in widely varying terms, making it substantially more challenging to conduct a comprehensive literature review.

Li et al. \cite{li_polygon-based_2014} point out the difficulties of extracting multilane roads from OSM that arise from each lane being represented as a separate linestring (a linestring is a geometry object representing a linear element such as a street edge following the Simple Features specification \cite{open_geospatial_consortium_simple_2023}). The authors propose a method to identify and merge face artifacts, which they call ``multilane polygons'', i.e., adjacent polygons covering a single street area resulting from mapping multiple street lanes, through a support vector machine (SVM) machine learning algorithm that uses five shape parameters as input. While this method does succeed in identifying face artifacts at multilane roads, it is only reproducible by users with advanced machine learning skills; furthermore, the method requires an input of manually classified training data, which adds substantial effort.

Fan et al. \cite{fan_polygon-based_2016} identify face artifacts, which they call ``non-urban block polygons'', as a data preprocessing issue in their study on feature matching between OSM and reference data. The authors use the SVM approach developed by Li et al. \cite{li_polygon-based_2014} mentioned above to identify face artifacts; they point out that the approach fails for smaller face artifacts at traffic junctions. 

In a similar use case but from a different field, Sanzana et al. \cite{sanzana_decomposition_2018} elaborate on the process of deriving hydrological response units for drainage network flow modeling and find that error correction is needed for  ``bad-shaped polygons''. The authors classify face artifacts, formed by roads and footpaths, as ``sliver polygons'', a subcategory of ``bad-shaped polygons''. The authors further propose a method of polygon decomposition into smaller ``well-shaped polygons'', which, however, cannot be conceptually transferred from hydrological to street networks.

Grippa et al. \cite{grippa_mapping_2018} develop a workflow for land use classification on an urban block level. The urban blocks are derived through polygonization of the OSM street network, which however brings about some artifacts, as the authors point out by distingushing between ``urban blocks'' (i.e., polygons correctly approximating urban blocks) and ``sliver polygons'' (i.e., polygons that do not approximate urban blocks). Sliver polygons are detected based on shape and size criteria, which are user-defined and not further specified or analyzed by the authors. To remove the sliver polygons from the data set, a semi-automated workflow, partially in PostGIS, is implemented. 

Ludwig et al. \cite{ludwig_mapping_2021} take up the approach by Grippa et al. \cite{grippa_mapping_2018}, as described above, within the context of urban green space identification from satellite imagery. So-called ``city blocks'', needed as units for the analysis, are derived through polygonization of the combined OSM networks of streets, railways, and waterways. To identify ``sliver polygons'', the authors take up the approach by Grippa et al. \cite{grippa_mapping_2018}, as described above, with an additional threshold criterion of minimum area.  

The study by Vybornova et al. \cite{vybornova_automated_2023} aims at identifying gaps in a network of bicycle infrastructure, which is defined as a subgraph of the OSM street network. For estimating network flow, the authors apply a shortest-path algorithm and point out that results are partially distorted by ``parallel edges'', i.e., the network edges at the boundaries of face artifacts. The study presents a network shortest path-based approach for the identification of parallel edges but no solution to effectively remove these from the network.

Lastly, a recent study by Shpuza \cite{shpuza_shape_2022} on the shape and size statistics of urban blocks describes elongated urban blocks that are delimited either by a street or another type of obstacle (e.g., a waterbody) and that contain no buildings, as ``edge blocks''. The author points out that edge blocks can be identified as outliers in a so-called ``shape matrix'' based on two geometrical parameters: relative distance and directional fragmentation. However, in this definition, edge blocks represent actual urban blocks rather than scattered parts of the street space.

The studies cited above demonstrate that face artifacts present a real data preprocessing challenge across a wide range of disciplines that work with street network data -- from hydrological flow modeling and satellite imagery-based land use classification to transportation planning and urban morphology. However, no previous study has so far tackled this issue systematically, which is also seen in the lack of a coherent terminology for this common problem. In addition, the solutions found in the literature for face artifact identification are either not reproducible, not transferrable to other use cases, or not automated. 

We conclude the literature review with a suggestion to consolidate the terminology. As described above, some studies refer to face artifacts as ``sliver polygons'' \cite{grippa_mapping_2018, sanzana_decomposition_2018, ludwig_mapping_2021}. However, these are two conceptually different terms since face artifacts arise as a consequence of a context-dependent redundancy of mapped line features, while sliver polygons stem from mismatching boundaries in vector overlays of polygon features \cite{goodchild_statistical_1978, fischer_using_1993, delafontaine_assessment_2009}. The terms ``multilane polygons'' \cite{li_polygon-based_2014} or ``parallel edges’’ \cite{vybornova_automated_2023} do not reflect other transportation geometries causing the issue (e.g., complex intersections), while ``bad-shaped polygons'' \cite{sanzana_decomposition_2018} is a vague term which does not refer back to the actual issue. We therefore suggest the term ``face artifact'', which is derived from graph theory and encodes both the \textit{origin} (i.e., network polygonization) and the \textit{erroneous nature} (i.e., an artifact in the context of street network polygonization) of the polygon.

\section{Data and Method}
\label{sec:method}

The method proposed in this article is a simple, computationally cheap procedure designed to capture both the artifacts resulting from dual carriageways and the artifacts resulting from complex intersections, including roundabouts. To develop, validate, and evaluate our method, we apply it to a data set of 131 cities randomly sampled from across the globe. In the subsections below, we first describe the retrieval of the data set, and then each of the methodological steps, which can be summarized as follows. First, the street network is polygonized to retrieve face polygons; then, the polygons are characterized according to their compactness and their area. Guided by the intuition that the distribution of urban block shape metrics has some universal properties, we combine both of these polygon characteristics to derive a \textit{face artifact index}, which allows us to distinguish between face artifacts and actual urban blocks. We validate our method for face artifact identification both visually and computationally. Lastly, we evaluate the outcomes for each compactness metric to determine which compactness metric the face artifact index should be based on. 

\subsection{Data: 131 cities' street networks from OpenStreetMap}

\begin{figure}[ht]
    \centering
    \includegraphics[width=\textwidth]{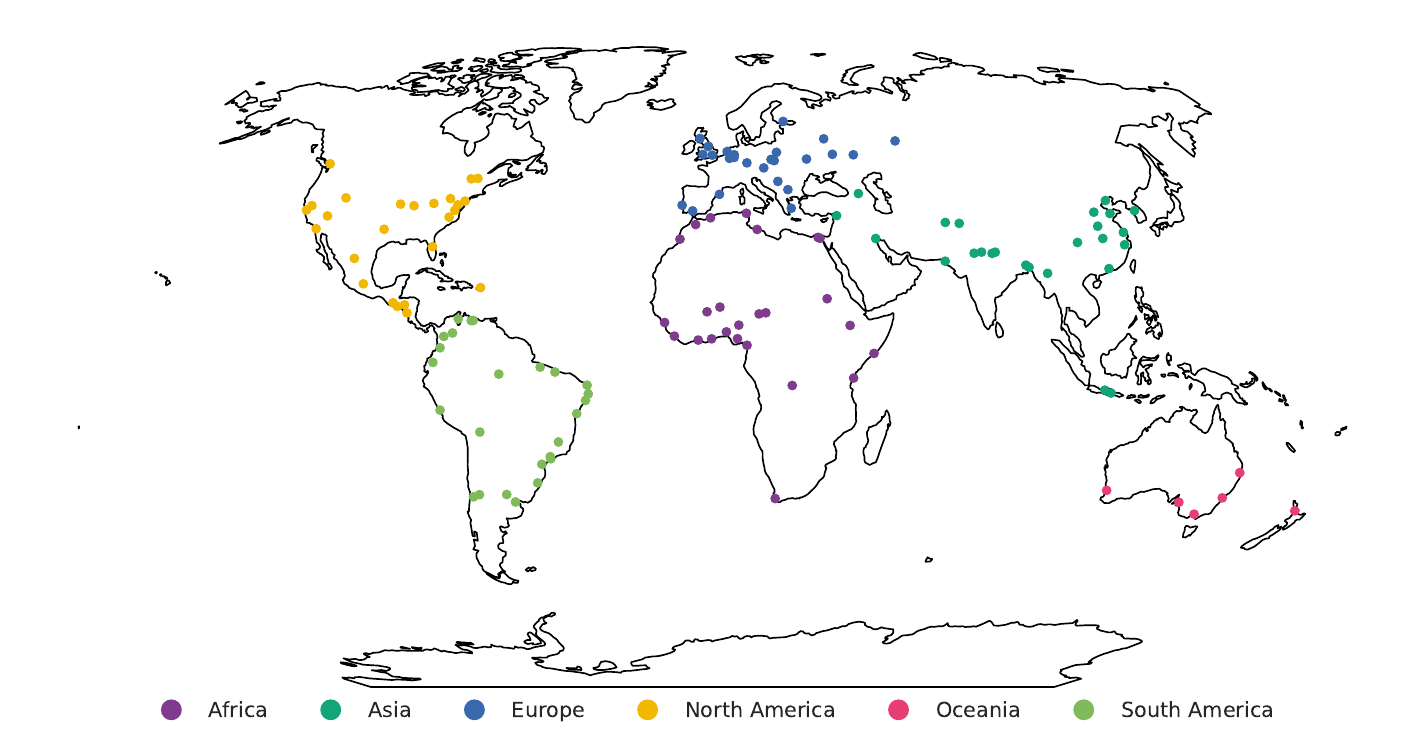}
    \caption{Spatial distribution of FUAs selected to be used within this study color-coded according to the (sub-)continent they lie on.}
    \label{fig:fua_map}
\end{figure}

The main goal of the method is to understand which components of a street network representation designed for other than morphological purposes are forming face artifacts and, hence, which edges need to be pre-processed prior to any analysis assuming morphological representation. Therefore, we need to retrieve data digitized with transportation in mind, like those available in OSM, which aims, among others, to include the level of detail required for GPS navigation. At the same time, the method shall be independent of the geographical context, and therefore, cannot depend on auxiliary data sources with limited availability (e.g., cadastral data). For these reasons, in this study we use OSM street network data for 131 cities sampled from six geographical areas (Africa, Asia, Europe, North America, Oceania, South America), covering all urbanized continents. To avoid ambiguity in the definition of a ``city'', we use the definition of functional urban areas (FUAs) released as part of the Global Human Settlement Layer (GHSL) \cite{schiavina_ghs-fua_2019, kemper_global_2021}. Given that smaller FUAs may contain only a low number of street edges, we further limit the selection to FUAs with at least 1 million inhabitants (as per GHSL data from 2015 released as part of the FUA layer). Out of this subset, we randomly select 25 FUAs per continent. For Oceania, which contains only six FUAs with more than 1 million inhabitants, all of these six are included in our sample. The geographical distribution of the thus selected FUAs is illustrated in Figure~\ref{fig:fua_map}, with a complete list available in the \nameref{sec:si}. The OSM data represents a subset of data sources with similar structure, and we expect the method to work equally on a wide range of other data sources with comparable structure (e.g.,~Overture Maps or many municipal data sources).
 
\subsection{Street network retrieval and generation of face polygons}

For each of the selected FUAs, we retrieve the street network data intersecting its boundary from OSM. We use only geometries with a \texttt{highway} tag and a filter based on OSM highway hierarchy (see \nameref{sec:si} for details on the custom filter used), ensuring we do not include sidewalks or service roads. Retrieved street geometries are further processed using \texttt{OSMnx} \cite{boeing_osmnx_2017} to derive a topologically correct undirected network, which is then polygonized to retrieve face polygons enclosed from all sides by planar network edges and representing either urban blocks or face artifacts.
 
\subsection{Compactness metrics}
The conceptual logic behind the face artifact identification is simple. Since street networks are relatively predictable geometries with characteristic patterns and a limited number of feasible configurations, we know that face artifacts are typically either large and highly elongated polygons (resulting from dual carriageways) or small, highly compact polygons of various rather compact shapes, often circular (e.g.~roundabouts) or triangular (e.g.~turning lanes on intersections); in other words, face artifacts have either a large area and low compactness, or small area irrespective of compactness. Therefore, as will be shown below, a shape index that captures the relationship between area and compactness of face artifacts should have similar values for both face artifact types. 

Compactness metrics, i.e., shape metrics that distinguish between compact and elongated polygons, are well known in literature both within urban morphology \cite{fleischmann_measuring_2021} and generic shape statistics \cite{altman_districting_1998}. Even though there is a large number of used metrics, they are often only minor variations of the same formula leading to a linear relationship between them, as in the case of \textit{isoperimetric quotient} in Altman \cite{altman_districting_1998} and \textit{form factor} in Sanzana et al.~\cite{sanzana_decomposition_2018}). Some studies use the same formula under a different name, e.g., the \textit{radii index} in \cite{flaherty_compactness_1992} and \textit{Schumm's shape index} in \cite{maceachren_compactness_1985}. For the purposes of our study, we select a subset of five compactness metrics (see Table \ref{tab:metrics}) from the extensive set of those available in the \texttt{esda} Python package \cite{rey_pysalesda_2021} belonging to the Python Spatial Analytics Library (PySAL) family \cite{rey_pysal_2022}. The five compactness metrics differ in their formulae, but all five are dimensionless ratios based on a comparison of an ideally compact shape (e.g., the equi-areal circle) and the actual shape of a polygon and are independent of the polygon's area and orientation in space. Note that due to the nature of their definitions, it is expected that some metrics should show high degree of correlation.

Some of the metrics used in the related literature were excluded for conceptual reasons, such as the \textit{convexity index} used by Sanzana et al. \cite{sanzana_decomposition_2018}, since long and narrow geometries can be as convex as short and wide ones, or \textit{fractal dimension} in Grippa et al. \cite{grippa_mapping_2018} since the difference between urban blocks and face artifacts does not reside in their fractal dimensions. Shpuza's indices based on space syntax theory \cite{shpuza_shape_2022} are conceptually within the scope of this research but have a level of complexity that does not fit our intention to propose a simple and computationally cheap heuristic due to the number of steps needed to derive a compactness metric for a polygon. Since the main benefit of this compactness index \cite{shpuza_shape_2022} is its ability to take into account cul-de-sacs in the interior of the polygon, which by definition cannot form a face artifact, it is omitted from this comparison. Grippa et al. \cite{grippa_mapping_2018} do not report formulae, so we can only assume how compactness metrics relative to a square and to a circle were measured. A compactness metric presented in Louf and Barthelemy \cite{louf_typology_2014} is equal to \textit{circular compactness} (also known as \textit{minimum bounding circle ratio}), first used by Reock \cite{reock_note_1961} and later by Frolov \cite{frolov_measuring_1975}, and likely also to the circular compactness as used by Grippa et al. \cite{grippa_mapping_2018}. Further, a metric presented as \textit{elongation} in Gil et al. \cite{gil_discovery_2011} is equal to the \textit{diameter ratio} presented earlier by Flaherty and Crumplin \cite{flaherty_compactness_1992}.

\begin{table}
    \centering
        \begin{tabular}{lccc}
            \hline
            Compactness metric $C_{i}$ & Formula & Reference & Face artifact index $F_{i}$\\
            \hline \\
            Circular compactness $C_{cc}$ & $\displaystyle\frac{a_{g}}{a_{mbc}}$ & \cite{reock_note_1961} & $F_{cc}$ \\\\
            Isoperimetric quotient $C_{ipq}$ & $\displaystyle\frac{4 \pi a_{g}}{p_{g}^2}$& \cite{altman_districting_1998} & $F_{ipq}$ \\\\
            Isoareal quotient $C_{iaq}$ & $\displaystyle\frac{2 \pi \sqrt{\frac{a_{g}}{\pi}}}{p_{g}}$& \cite{altman_districting_1998} & $F_{iaq}$ \\\\
            Radii ratio $C_{rr}$ & $\displaystyle\frac{\sqrt{\frac{a_{g}}{\pi}}}{r_{mbc}}$& \cite{flaherty_compactness_1992} & $F_{rr}$ \\\\
            Diameter ratio $C_{dr}$ & $\displaystyle\frac{w_{mrr}}{l_{mrr}}$ & \cite{flaherty_compactness_1992} & $F_{dr}$
        \end{tabular}
     \caption{A selection of compactness metrics $C_{i}$ tested within this study, where
         $a_{g}$ = area of a polygon;
         $a_{mbc}$ = area of minimum bounding circle of a polygon;
         $p_{g}$ = perimeter of a polygon;
         $r_{mbc}$ = radius of a minimum bounding circle of a polygon;
         $w_{mrr}$ = width of a minimum rotated rectangle of a polygon, where width is equal to the smaller dimension of the rectangle;
         $l_{mrr}$ = length of a minimum rotated rectangle of a polygon, where length is equal to the larger dimension of the rectangle.}
     \label{tab:metrics}
\end{table}

\subsection{Proposed heuristic: Face artifact index}

For each face polygon $p$ in every FUA, we compute its area and the compactness metric  $C{i,p}$ (where $i$ denotes one of the five selected compactness metrics: circular compactness, isoperimetric quotient, isoareal quotient, radii ratio, and diameter ratio). Then, for each compactness metric we derive a corresponding \textit{face artifact index} $F_{i,p}$, that captures both a polygon's compactness and area:

\begin{equation}
\label{eq:1}
F_{i,p}=\log(C_{i,p}*a_{p})
\end{equation}

\noindent The multiplication captures the relationship between the compactness and area, while logarithmic scaling smoothens out the heavy-tailed distribution of the index caused by outliers (mostly geometries with particularly large areas). Next, we observe the frequency distributions $\phi_{i}=\phi(F_{i})$ of the five face artifact indices for each FUA. For the majority of the analyzed FUAs and compactness metrics, the $\phi_{i}$ reveal a common feature of showing at least two well-pronounced peaks, i.e., two prominent local maxima, as will be shown in Figure~\ref{fig:phis}. 

Through visual analysis, we estimate that these peaks represent two different types of polygons: most polygons within the first (leftmost) peak can be attributed to face artifacts in the street network, whereas most polygons within the second (rightmost) peak represent actual urban blocks. Therefore, for FUAs that show a pronounced two-peak pattern in their $\phi(F_{i})$ distribution, we define the \textit{face artifact index threshold $T_{i}$} as the value of $F_{i}$ corresponding to the valley, i.e., the local minimum \textit{between} the two peaks in the distribution $\phi_{i}$. To find the value of $T_{i}$ for a given FUA, we approximate $\phi_{i}$ with a Gaussian kernel density estimation (see \nameref{sec:si} for specifications). Then, in order to account for different urban forms that result in varying shapes of $\phi$ for different FUAs (some FUAs' $\phi_{i}$ distributions have more than two peaks), we formulate two conditions: $(1)$ the face artifact index distribution has \textit{at least} two peaks; $(2)$ the face artifact index distribution has \textit{at least} one valley. If both these conditions are fulfilled for a given FUA and a face artifact index $F_{i}$, we define the face artifact index threshold $T_{i}$ as the value of $F_{i}$ at the location of the first valley that lies between two peaks, one of which is the highest one:

\begin{equation}
\begin{split}
T_{i} = F_{i} | \phi'(F_{i}) = 0~\land \\
\phi''(F_{i}) > 0~\land \\
\exists~G_{i}<F_{i} | ( \phi'(G_{i})=0 \land \phi''(G_{i})<0)~\land \\
\exists~H_{i}>F_{i} | ( \phi'(H_{i})=0 \land \phi''(H_{i})<0)~\land \\
\arg\max(\phi) \in [G_{i}, H_{i}] 
\end{split}
\end{equation}

\noindent We postulate that once we have computed, for a given FUA and a selected compactness metric, both $F_{i,p}$ and $T_{i}$, the face polygons can be easily classified into artifacts and urban blocks. Polygons with a face artifact index below the threshold ($F_{i,p}<T_{i}$) will most likely be face artifacts; polygons with a face artifact index at or above the threshold ($F_{i,p} \geq T_{i}$) will most likely be urban blocks. 

In the next steps, we describe the methods used to check the validity of the proposed heuristic and to evaluate which compactness metric is best used for face artifact detection.

\subsection{Validation: Visual assessment and overlay with OSM building data}

To validate our proposed heuristic and to assess whether the face artifact identification actually yields satisfactory results, we conduct two validation steps. First, we validate our method visually by generating plots of all face artifacts for each FUA where a face artifact index threshold could be identified. All plots can be found in the project repository archive available from \url{doi.org/10.5281/zenodo.8300729}; Figure~\ref{fig:birdview} shows four examples. 

Next, we validate our method computationally using OSM building data. By our own definition of face artifacts, these are part of the street network surface of a city and, therefore, should not contain any buildings -- with minor exceptions, for example, when a bus stop in the middle of a road is mapped out as a separate building, or when a road leads underneath a building. Thus, from the perspective of face artifact identification, we can use OSM building data to identify ``false positives'' -- polygons that have been classified as face artifacts by our method but contain (or overlap with) building footprints. Note that this is the only corner of the confusion matrix for our face artifact identification heuristic that we are able to estimate computationally in the absence of ground truth. Whether ``negatives'' (polygons classified as urban blocks) are true (actual urban blocks) or false (wrongly classified face artifacts) cannot be evaluated through building footprints, since urban blocks may or may not contain buildings. The same goes for ``true positives'' (correctly identified face artifacts): we cannot estimate their number computationally since the absence of building footprints is a necessary but not a sufficient condition for a face polygon to be classified as face artifact. 

Since the completeness of OSM building data is an ongoing object of study \cite{biljecki_quality_2023, oostwegel_automatic_2023} and can vary greatly depending on the region, we first download and visually inspect OSM building data sets for all FUAs. We discard FUAs with low coverage of mapped buildings and keep only FUAs with sufficiently complete OSM building data. Then, for each FUA that has both a detected face artifact index threshold and sufficient building data, we compute the overlap of face polygons and building footprint polygons. We then compute the percentage of false positives, i.e., the percentage of face artifacts that contain up to $X$ square meters of building footprints, letting $X$ vary between $0$ and $100~m^{2}$ to account for building overlap exceptions described above (see Figure~\ref{fig:valid}). 

\subsection{Evaluation of compactness metrics}

Next, we aim to select one out of the five analyzed compactness metrics, based on which the face artifact index threshold shall be computed. To this end, we compare the performance of compactness metrics $C_{i}$ listed in table \ref{tab:metrics} based on their ability to detect face artifacts and on their computational efficiency. An overview of all evaluation steps is shown in Figure~\ref{fig:eval}.

The ability to detect a threshold $T_{i}$ that divides face polygons into artifacts and urban blocks is the most critical aspect of the proposed heuristic. The first evaluation step therefore captures the percentage of successfully detected $T_{i}$ for each compactness metric $C_{i}$. It is assumed that the variation in distributions among metrics will result in differences reflecting the quality of each metric. 

The second evaluation computes the peak prominence of the face artifact index threshold, i.e., the vertical distance between the valley (corresponding to $T_{i}$) and the peak to its left in the distribution $\phi_{i}$. This metric captures how easy it is to derive the threshold from the distribution itself, depending on which compactness metric $C_{i}$ the computation is based on. 

Next, we evaluate the computational efficiency, as we believe that the lower the usage barriers of a method, the higher its value for the scientific community. Computational efficiency is one potential usage barrier, excluding researchers with limited access to high-performance computing infrastructure. We therefore benchmark the single-threaded performance of each metric on a random sample of $10000$ polygons taken from the pool of all FUAs. The implementation of all compactness metrics in the \texttt{esda} package relies on the vectorized geometry engine of a \texttt{shapely 2} \cite{gillies_shapely_2023} wrapping around the \texttt{GEOS} library \cite{geos_contributors_geos_2023}, ensuring minimal processing overhead. The benchmark is run on an Intel Xeon W-2245 3.9GHz CPU, repeating the same measurement 100 times.

The fourth and last evaluation step is based on the validation with OSM building data and looks at the percentage of ``false positives'', i.e., the percentage of polygons that have been identified as face polygons but contain buildings. The false positives have been computed only for those FUAs where OSM building data are sufficiently complete; therefore this evaluation step has fewer data points.

\section{Results}
\label{sec:results}

Below, we summarize the results for each of the methodological steps, described in detail in Section \ref{sec:method}). Detailed results can be found in the project repository archive \url{doi.org/10.5281/zenodo.8300730}.

Using the sample of 131 FUAs, we were able to retrieve a total of $3440838$ face polygons, with ~ $25\%$ in North America, $23\%$ in South America, $16\%$ in Europe, $14\%$ in Africa and $13\%$ and $7\%$ in Asia and Oceania respectively, and a mean number of over $26000$ polygons per FUA. That leaves us with a large amount of data ensuring a certain level of robustness and sampling bias reduction.

\subsection{Compactness metrics}

\begin{figure}[htbp]
    \centering
    \includegraphics[width=\linewidth]{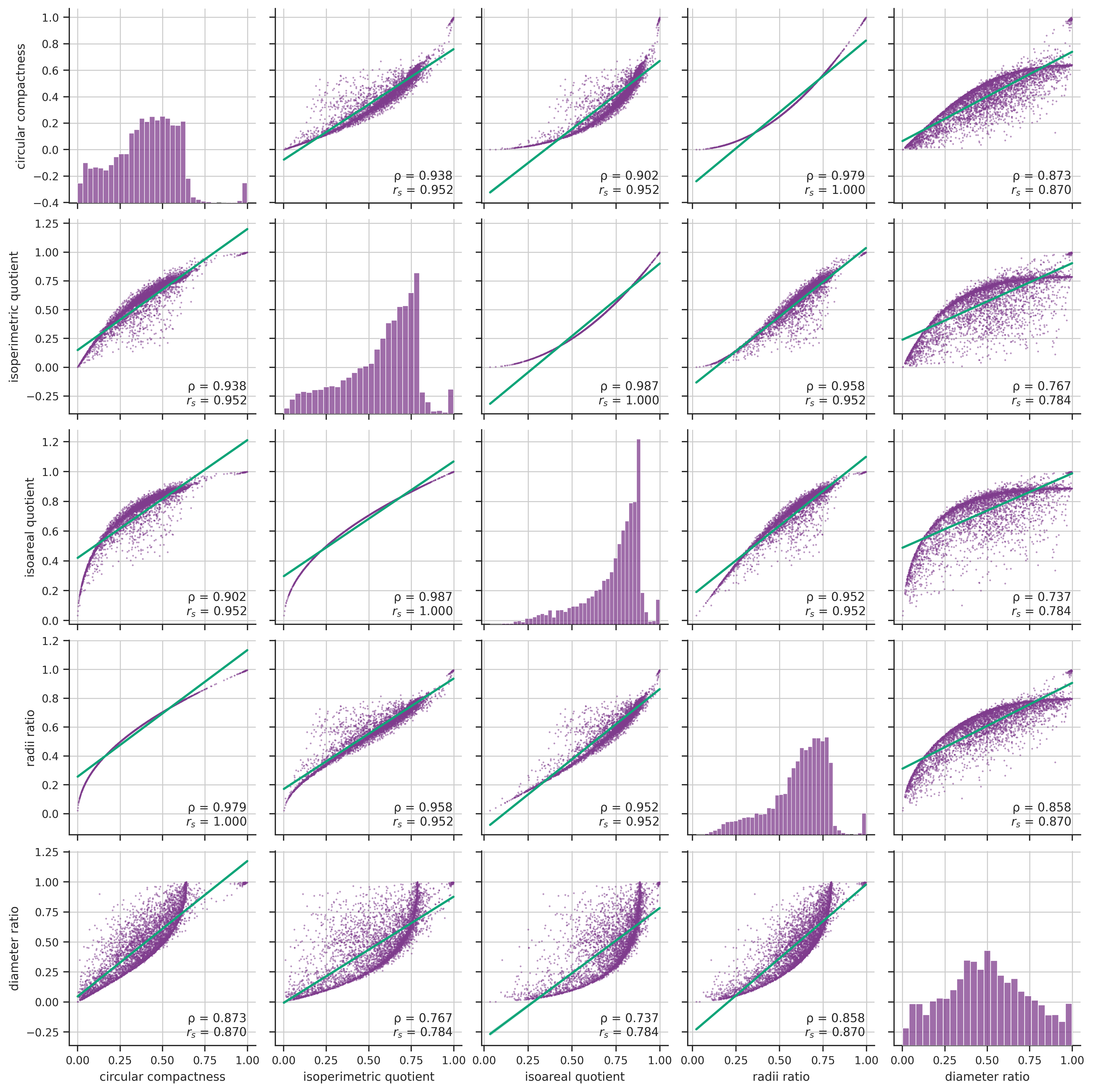}
    \caption{Pairwise scatter plots and histograms of five shape metrics for a combined sample of all 131 FUAs. The diagonal plots show the distribution of each metric, while the off-diagonal plots show the correlation between the two metrics. Each scatter plot contains a value of Pearson correlation coefficient ($\rho$) and of Spearman's rank correlation coefficient ($r_{s}$) for a specific pair.}
    \label{fig:pairwise}
\end{figure}

As expected, there is a strong relationship between the individual compactness metrics as shown in Figure~\ref{fig:pairwise}. The Pearson correlation coefficient ranges from $0.735$ between the \textit{isoareal quotient} and the \textit{diameter ratio} to $0.987$ between the \textit{isoperimetric quotient} and the \textit{isoareal quotient}. The latter is due to a non-linear relationship between the two metrics, evident both in Figure~\ref{fig:pairwise} and in Spearman's rank correlation coefficient of $1.0$. Other pairs of metrics clearly capture similar shape characteristics but do not have a direct (linear or non-linear relationship). We can therefore assume that all five compactness metrics should be able to detect face artifacts in a similar manner, though likely with a different performance.

\begin{figure}[htbp]
    \centering
    \includegraphics[width=\linewidth]{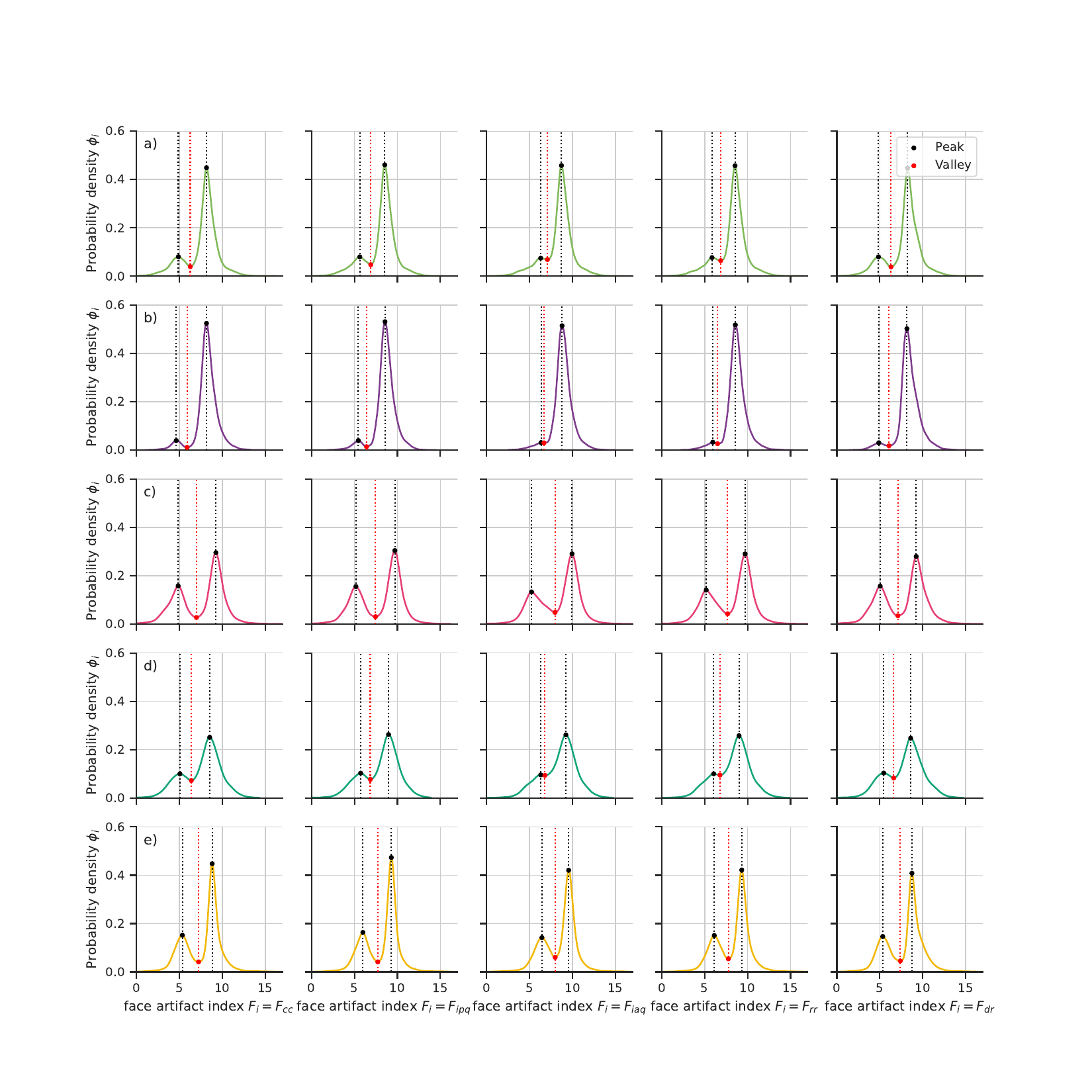}
    \caption{Face artifact index distributions for different compactness metrics and cities. In the columns, from left to right: circular compactness; isoperimetric quotient; isoareal quotient; radii ratio; diameter ratio. In the rows, from top to bottom: a) Cochabamba (Bolivia); b) Douala (Cameroon); c) Sydney (Australia); d) Tbilisi (Georgia); e) Montreal (Canada). Peaks are highlighted with black dots; valleys with red dots. The dashed red vertical line shows the position of the identified face artifact index threshold $T_{i}$ for the given compactness metric and city.}
    \label{fig:phis}
\end{figure}

\noindent This assumption is confirmed once we take a look at the distributions $\phi_{i}$, as shown in Figure~\ref{fig:phis} for five different cities: all face artifact indices $F_{i}$ appear to yield similar two-peak patterns of $\phi_{i}$, though the peak prominence and exact curve shape vary depending on the compactness metric used.

\subsection{Validation}
\begin{figure}[htbp]
    \centering
    \includegraphics[width=0.9\textwidth]{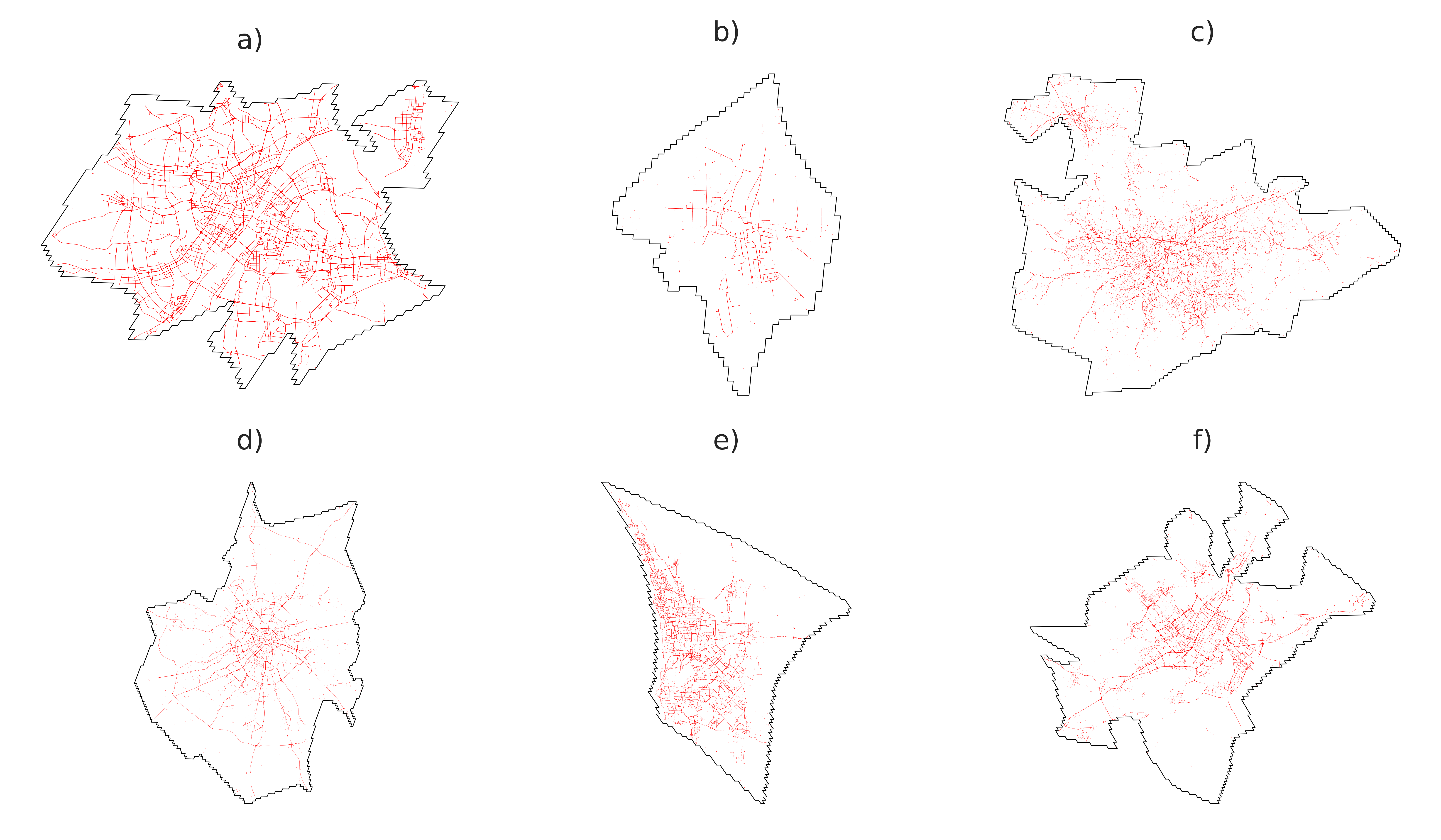}
    \caption{Birdview plots of detected face artifacts (red polygons) within each FUA border (black lines) for one city per each continent, clockwise from top left: Wuhan (China, Asia), Khartoum (Sudan, Africa), São Paulo (Brazil, South America), Moscow (Russia, Europe), Perth (Australia, Oceania), and Montreal (Canada, North America). \url{doi.org/10.5281/zenodo.8300730}.}
    \label{fig:birdview}
\end{figure}

The visual validation step provides a first insight into the success of our method. We generate separate plots of face artifacts for each FUA  where a face artifact threshold could be identified (see \url{doi.org/10.5281/zenodo.8300730}) and note that almost all of them clearly align with street network patterns, as can be seen in Figure~\ref{fig:birdview}. The only clear outlier is Mogadishu (Somalia), where entire areas of the city have been wrongly marked as face artifacts by our method -- notably the districts of Shibis and Karan, which are characterized by a dense rectangular street pattern with a typical grid cell length of around $30~m$. The corresponding face artifact index distribution suggests that the threshold should be set to the first valley (and not the one to the left of the highest peak) in order to obtain better results. 

\begin{figure}[htbp]
    \centering
    \includegraphics[width=\textwidth]{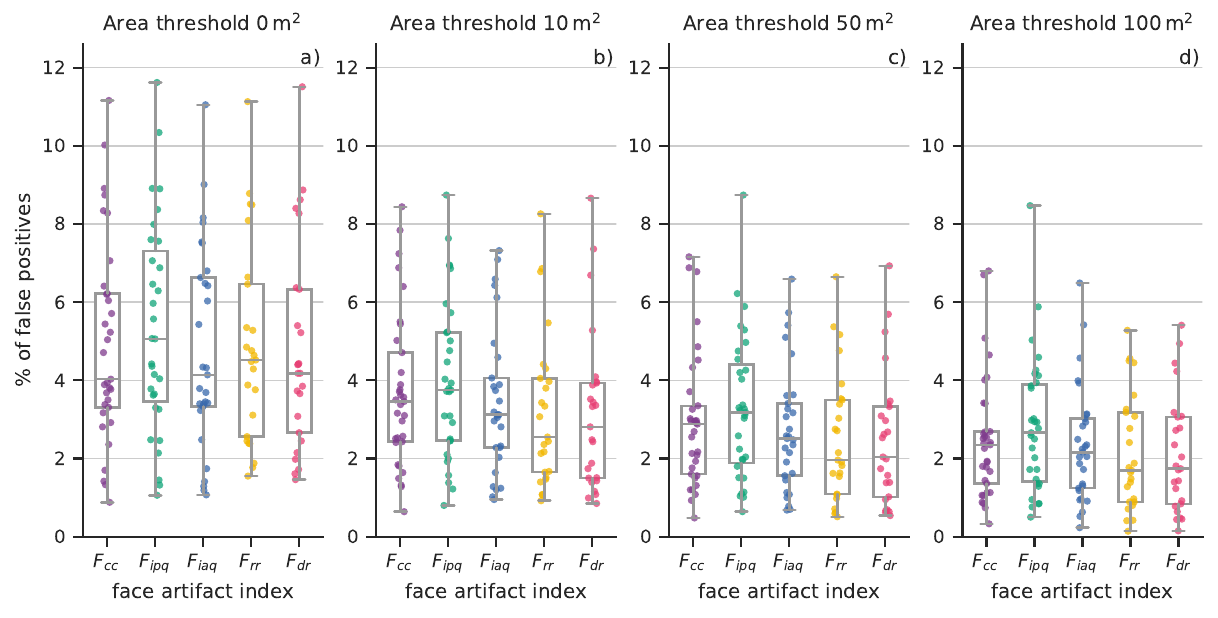}
    \caption{Validating the heuristic with OSM building data}
    \label{fig:valid}
\end{figure}

The computational validation step also shows satisfactory results. As seen in panel b) of Figure~\ref{fig:valid}, for the threshold based on circular compactness and with an area threshold of $10~m^{2}$, for most FUAs the rate of false positives lies between $2.5$ and $4.5\%$. In other words, for most FUAs, only $2.5-4.5\%$ of detected face artifacts overlap with OSM building data, while the overwhelming majority of face artifacts ($95.5-97.5\%$) overlap with no building footprints or with a negligible extent ($ \leq 10~m^{2})$ of building footprints. Figure~\ref{fig:valid} further shows that we get comparable results for face artifact index thresholds based on other compactness metrics.

\begin{figure}[htbp]
    \centering
    \includegraphics[width=0.9\textwidth]{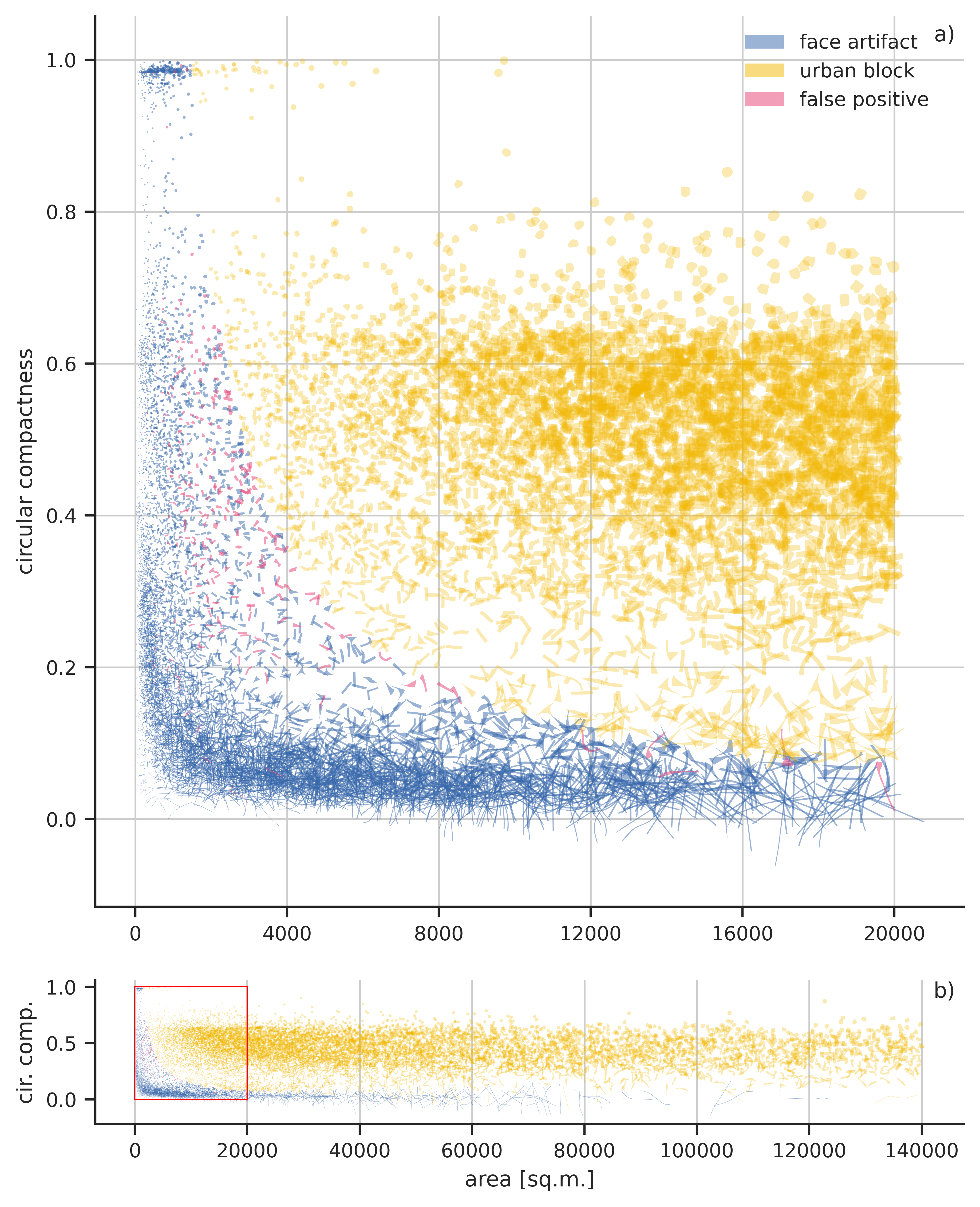}
    \caption{A scatter plot of individual polygons based on their area and circular compactness, with individual groups highlighted based on the result of the face artifact detection method and its validation. The panel (a) reflects a zoomed-in version of the bottom left corner of the full figure in panel (b).}
    \label{fig:banana}
\end{figure}

\noindent Figure~\ref{fig:banana} shows the relationship between polygon compactness and area for all face polygons within the FUA of Raleigh (USA). The plot distinguishes between identified urban blocks (blue), identified face artifacts with no building overlap (yellow) and identified face artifacts with building overlap, i.e., false positives (pink). The distribution visually indicates two large clusters, one representing face artifacts and the other one capturing urban blocks. However, it is to be noted that while the area where the threshold lies is sparsely occupied, there is no clear-cut boundary between the polygon groups, at least not visually.

\subsection{Evaluation}
After confirming that our heuristic is indeed able to identify face artifacts, we need to decide which of the five compactness metrics $C_{i}$ to settle on. To evaluate the compactness metrics, we look at four aspects: percentage of FUAs with an identified threshold; computational efficiency; peak prominence; and percentage of false positives (see Section \ref{sec:method} for details).

As illustrated in panel a) of Figure~\ref{fig:valid}, the face artifact index based on circular compactness could identify the greatest number of thresholds, namely in 117 out of 131 FUAs (89\%). The least number of thresholds was identified by the face artifact indices based on isoareal quotient and radii ratio, with 77 (58\%) and 83 (63\%) identified FUA thresholds, respectively. The face artifact indices based on isoperimetric quotient and diameter ratio both detected a threshold in 109 cases (83\%), though not the same set. It is worth noting that even the face artifact index based on circular compactness does not capture all FUA thresholds captured by the other indices. For example, the threshold for Ibadan (Nigeria), not found by the former, is indeed identified by three other face artifact indices (based on isoperimetric quotient, radii ratio, and isoareal quotient). Furthermore, the face artifact index based on diameter ratio allowed the identification of thresholds in three additional cases: Dhaka (Bangladesh), Jombang (Indonesia), and Recife (Brazil). Neither one of the compactness metrics, therefore, seems to be universally superior to the others. However, for the particular sample of 131 FUAs used in this study, circular compactness wins in absolute numbers of identified thresholds.

Analyzing the peak prominence for each of the shape metrics tells a similar story as the evaluation of detected thresholds above, but with a few differences. The metric with the highest peak prominence is the face artifact index based on the isoperimetric quotient, closely followed by the circular compactness, both with a mean of $0.32$. The face artifact index based on the isoareal quotient shows the least prominent peaks, with a mean value of $0.28$. However, the overall differences in peak prominence do not appear to be large, with all of the distributions following roughly the same shape leading to similar distances between their peaks and valleys; no one metric can therefore be singled out in contrast to the others.

\begin{figure}[htbp]
    \centering
    \includegraphics[width=\textwidth]{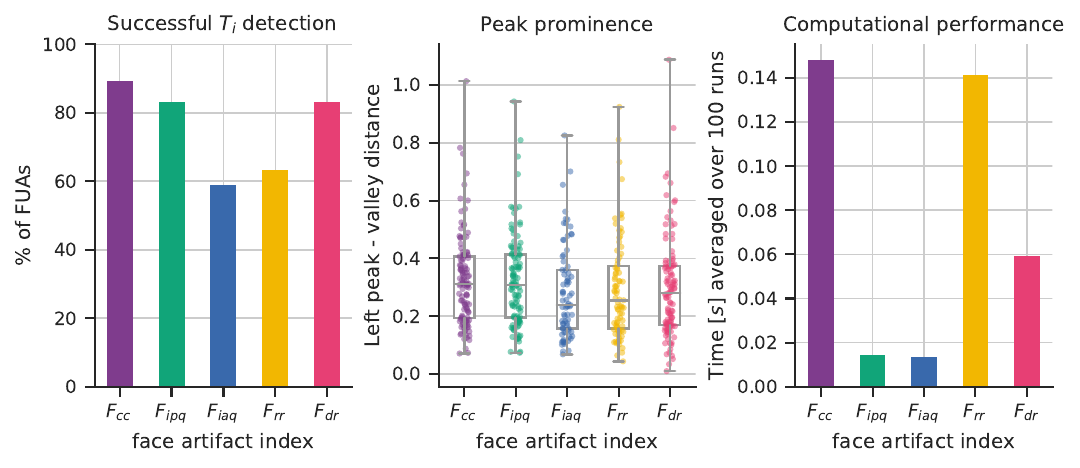}
    \caption{Evaluating compactness metrics}
    \label{fig:eval}
\end{figure}

Results from the computational performance benchmark, tested on $10000$ randomly sampled polygons, are plotted in the third panel of Figure~\ref{fig:eval}. Our heuristic is indeed computationally cheap (as intended) since even the slowest metric takes only an average of $147~ms$. This translates into a nearly negligible approximate processing time of $400~ms$ per average-sized FUA. When comparing the face artifact indices against each other, there are notable differences in computational speed: $F_{iaq}$ and $F_{ipq}$, the two fastest metrics, are more than $10$ times faster than $F_{cc}$ and $F_{rr}$, probably due to the fact that the latter requires computation of minimum bounding circles. From the perspective of computational performance, $F_{cc}$ seems to be the worst available option in comparison to the other face artifact indices; however, given how computationally cheap the heuristic is even for $F_{cc}$, this evaluation step seems to be relevant only for much larger data sets than the one used in this study. When using the GEOS-based implementation either directly in C++ or through efficient bindings (\texttt{shapely} in Python, \texttt{sf} or \texttt{terra} in R, or \texttt{PostGIS} in \texttt{PostgreSQL}) for a data set in the order of the one used in this study, it should therefore not matter which shape metric is used from a perspective of computational performance.

We compute the percentage of false positives for four different area threshold values of $X \epsilon \{0,10,50,100\} [m^{2}]$ and compare the  results for all five face artifact indices $F_{i}$. The overall percentage of false positives is decreasing with an increasing $X$ (as was to be expected), but is reasonably low even for the strict requirement of $X=0$, i.e., for the case in which \textit{any} amount of overlap with the OSM building data set leads to an identified face polygon to be classified as ``false positive''. Similarly to the previous evaluation steps, comparing the percentage of false positives by face artifact index $F_{i}$ does not indicate any clear winner. 

\section{Discussion}
\label{sec:discussion}

Unlike previous attempts based on complex machine learning algorithms \cite{li_polygon-based_2014, fan_polygon-based_2016}, neither of which is easily reproducible, the method proposed in this study is a simple heuristic based on properties of individual geometries. Our method is computationally cheap, transferrable, and easy to reproduce and replicate. By leveraging the characteristic patterns in urban street networks and their geometric representation, the face artifact index proposed in this study manages to capture both types of face artifacts, i.e., the elongated polygons between dual carriageways as well as small polygons of various shapes resulting from complex intersections, with a single metric and a single threshold.

There is no clear indication as to which compactness metric the face artifact index should be best based on -- rather, this will depend on the use case and on the data set available. The performance of $F_{iaq}$ and $F_{ipq}$ suggest that they should not be used but the decision on the remaining metrics is not that clear. For the data set we worked with in this study, where the primary goal was to identify face artifact index thresholds in as many as possible of 131 cities, the best choice is $F_{cc}$, i.e., the face artifact index based on circular compactness. However, for several cities in our data set, $F_{dr}$ would have been the better choice. Thus, we maintain the formula for $F_{i,p}$ introduced in Equation~\ref{eq:1} in general terms, where the compactness metric $C_{i}$ has to be chosen at the analyst's discretion.

However, there are also some shortcomings of our method. In 11\% of cases, the $\phi_{i}$ distribution does not follow a two-peak pattern and the threshold heuristic is not applicable. While that may mean that there are no face artifacts in the whole FUA, that is not very likely. We have observed varying percentages of artifacts within FUAs depending on their geographical location, reflecting differences in cultural background and the way it is translated into the urban structure. However, there are likely at least some polygons that can be considered artifacts. Cases with a low percentage of artifacts will remain without a detected threshold. It is possible that a threshold derived at a geographically close location will be applicable in these cases, but we recommend proceeding with caution when applying an externally computed threshold to a study area. 

A low number of artifacts can be caused not only by the lack of physical objects of a transportation origin, but also by the quality of data. OSM, which was used to retrieve all the street network geometries, has a very high coverage of the world when it comes to the geometries with a ``highway'' tag (consisting not only of highways but also other drivable or walkable linear elements like roads and paths). Unfortunately, the same cannot be said about the quality of the data. When there is a complex intersection, roundabout, or dual carriageway on the ground, its representation in OSM data may significantly differ from a detailed drawing in one case and only a single node in the other. Naturally, the latter case does not produce any face artifacts, while the former can create several face artifacts in one single location.

 \begin{figure}[ht]
    \centering
    \includegraphics[width=0.7\textwidth]{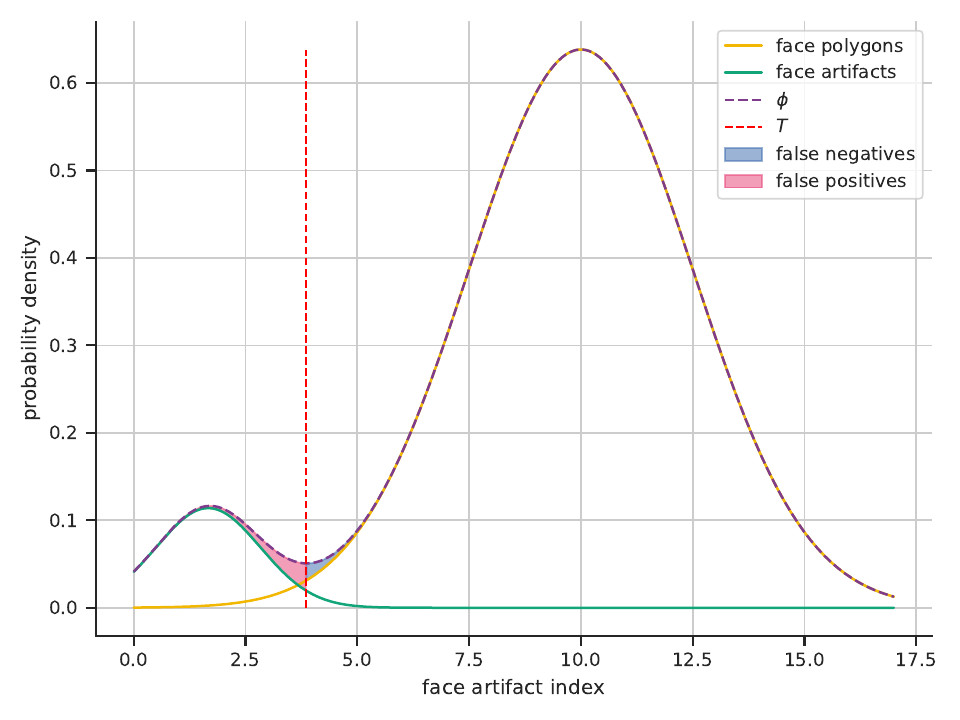}
    \caption{Conceptual illustration of the composition of $\phi$ consisting of two overlapping Gaussian distributions and resulting identification of false positives and false negatives when relying on $T$.}
    \label{fig:concept}
\end{figure}
 
The nature of the distribution with two local maxima required for the detection of $T_{i}$ comes with another limitation. In the ideal situation, we can conceptually split the distribution into two Gaussian distributions, one capturing the face artifact index of artifact poylgons and the other capturing the face artifact index of actual urban block polygons, as illustrated in Figure~\ref{fig:concept}. Their two tails overlap, leading to both false positive and false negative classification of polygons as artifacts (also illustrated by Figure~\ref{fig:banana}). We have seen in Figure~\ref{fig:eval} that the issue of false positives is relatively low, the median value ranging from 2-4\% depending on the selected compactness metric, but there is no straightforward way of quantifying false negatives. However, if we accept the conceptualization based on two normal distributions, we can assume that the number of false negatives is even lower than the number of false positives.

 \begin{figure}[ht]
    \centering
    \includegraphics[width=\textwidth]{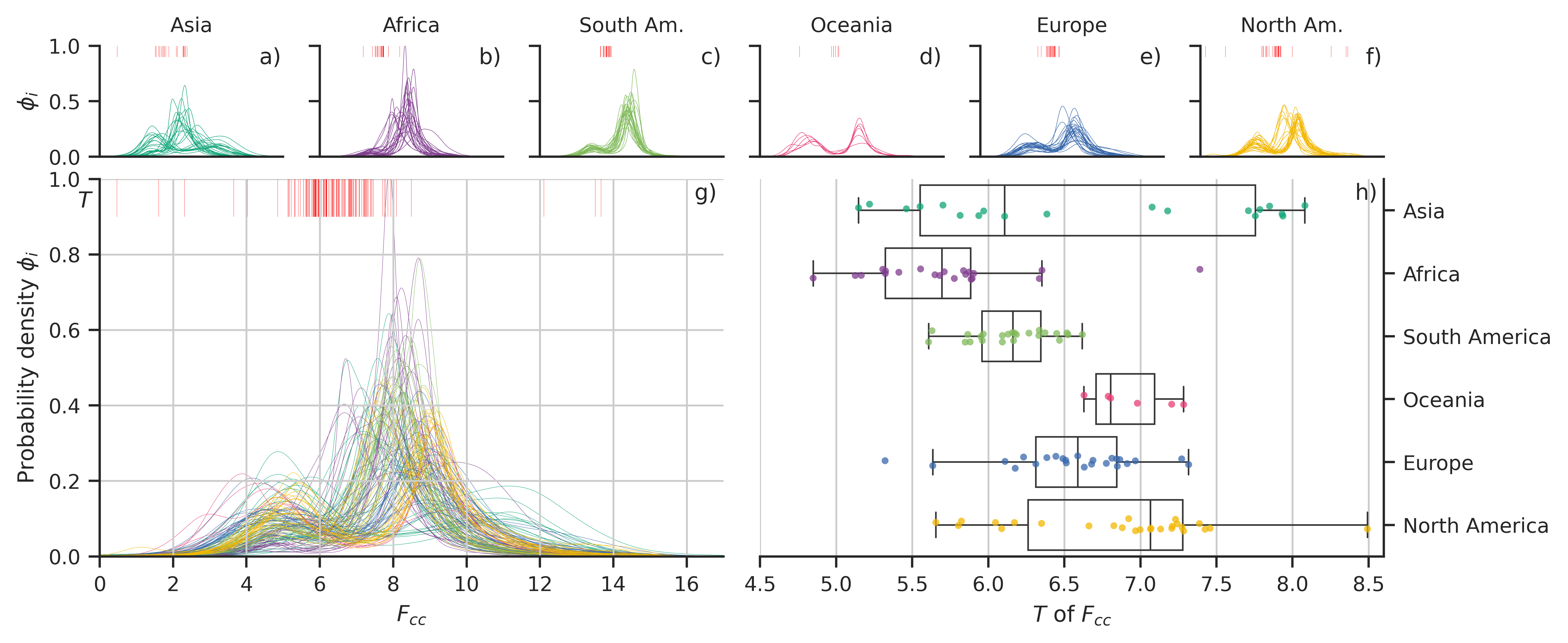}
    \caption{KDEs of the face artifact index in individual FUAs grouped by the (sub)continent (a-f); a combined plot of all the distributions (g) and a box plot showcasing the differences in the value of $T_{cc}$ across geographical locations (h). Each distribution plot also includes a rug plot of locations of individual $T_{cc}$ values in red.}
    \label{fig:geog_differences}
\end{figure}

Due to the design of urban road systems, contiguous artifacts often capture roads of a higher hierarchy, and their presence (or a lack thereof) reflects differences between design principles applied to cities in different parts of the world. These differences are clearly visible in the location of thresholds and the overall distributions of the face artifact index, as illustrated in Figure~\ref{fig:geog_differences}. While many distributions across all (sub)continents show a similar bimodal pattern, there are notable differences in the peak prominence and locations of the peaks and valleys. The most prominent thresholds are found in North American and Australian cities, most likely due to the prevalent car-oriented urban design that generates a large number of face artifacts as a byproduct. Asian cities seem to be divided into two subgroups, where the first one tends to have the first peak more prominent than the second, contrary to all the other cases. These distributions reflect street networks of large Chinese cities like Wuhan (illustrated in Figure~\ref{fig:birdview}a), Guangzhou or Chongqing, characterised by large-scale expansions with a highly car-oriented transport model in the last 50 years. These expansions lead to a large number of multiple-lane roads and complex intersections, causing an amount of face artifacts that exceeds the number of urban blocks. On the other side of this spectrum is the lowest peak prominence visible in African cities, likely due to a combination of less developed motorized transportation infrastructure compared to Chinese or North American cities, and a less precise mapping of such elements in  OpenStreetMap. These differences also show in the ability to detect $T_{cc}$ values in different regions, with Asia and Africa reaching the lowest value of 76\%, while in the rest of the world the method succeeds in detecting $T_{cc}$ in 96\% (South America, Europe) or even 100\% (North America, Oceania) of cases.

When focusing on the absolute values of $T_{cc}$, illustrated in Figure~\ref{fig:geog_differences}h, we can observe clearly differing trends for different continents. The exception, once again, are Chinese cities, which are more similar to North American cities than to other Asian ones. Given that the compactness component of $F_{cc}$ takes on values between 0 and 1, the main reason for these shifts is the size of the artifacts, where a higher $T_{cc}$ reflects generally larger face artifacts. As illustrated in Figure~\ref{fig:banana}, those are very elongated polygons, formed by large segments of dual carriageways. A lower $T_{cc}$ is often linked to a smaller number of highways, as we observe in the our sample of cities in Africa, South America, and Asia (except China). In contrast, places with strongly developed road infrastructure, such as Oceania (Australia), North America, and China, report higher $T_{cc}$. However, even while there are differences, the overall distributions are shifted only in a few percent of the extent, as shown Figure~\ref{fig:banana}g.

We call for future work to focus on the second part of the pursuit towards automatized simplification and transformation of street networks. Now that face artifacts can be computationally identified, the next step will be to develop an algorithm that can eliminate and appropriately replace face artifacts. 

In addition, the face artifact detection results are worth exploring on their own as they uncover structural differences between selected FUAs. The value of $T$ differs across locations; the shape of $\phi$ does not always resemble a bimodal distribution as shown in Figure~\ref{fig:concept}, and even when it does, the locations of peaks and the spread of distributions differ. The more profound exploration of regularities and irregularities in these results in connection with the geographical location may uncover higher-level principles guiding the development of global urban networks.

Every geospatial dataset is collected with a specific purpose in mind, which affects what is being captured, in which detail, where, and how. Nevertheless, the original purpose may not be the only one for which the dataset is potentially useful. Other use cases may arise, and with them, a need to adapt the data to fit the new purpose. In those cases, we need methods and processes allowing us to detect what needs to be changed and (ideally) to automatically adapt the input to the desired output. In the case of street networks captured with transportation in mind, the transformation process cannot be fully automated yet, but we believe that the method presented in this article provides an important step toward this goal.

\subsection*{Data and code availability}

The whole method is encapsulated in a series of Jupyter notebooks executed in a containerized environment \texttt{darribas/gds\_py:9.0} \cite{arribas-bel_gds_env_2023}, ensuring full reproducibility. All components of the work rely on open source software and open data, with the resulting code and data being openly available at \url{https://github.com/martinfleis/urban-block-artifacts} and archived at \url{doi.org/10.5281/zenodo.8300730}. The face artifact functionality has been contributed to the open source package \texttt{momepy} \cite{fleischmann_momepy_2019}, focusing on the analysis of urban morphology.

\section*{Acknowledgments}
The authors declare no conflicts of interest. M.F. kindly acknowledges funding by the UK’s Economic and Social Research Council through the project ``Learning an urban grammar from satellite data through AI'', project reference \texttt{ES/T005238/1} covering the initial exploration of the issue.

\bibliographystyle{josisacm}
\bibliography{references}

\newpage
\section*{Supplementary Material}
\label{sec:si}

\subsection{List of analyzed FUAs}
25 sample FUAs in \textbf{Africa}: Abidjan (CIV); Abuja (NGA); Accra (GHA); Addis Ababa (ETH); Agadir (MAR); Al-Zaqaziq (EGY); Amaigbo (NGA); Cape Town (ZAF); Conakry (GIN); Douala (CMR); Fez (MAR); Ibadan (NGA); Kananga (COD); Khartoum (SDN); Maiduguri (NGA); Mogadishu (SOM); Mombasa (KEN); Monrovia (LBR); N'Djamena (TCD); Niamey (NER); Oran (DZA); Ouagadougou (BFA); Tanta (EGY); Tripoli (LBY); Tunis (TUN). 

\smallskip

\noindent 25 sample FUAs in \textbf{Asia}: Abbottabad (PAK); Agra (IND); Aleppo (SYR); Basra (IRQ); Chongqing (CHN); Comilla (BGD); Dhaka (BGD); Gonda (IND); Guangzhou (CHN); Jaipur (IND); Jombang (IDN); Kabul (AFG); Karachi (PAK); Lucknow (IND); Luohe (CHN); Mandalay (MMR); Nantong (CHN); Qinhuangdao (CHN); Semarang (IDN); Seoul (KOR); Tbilisi (GEO); Weifang (CHN); Wuhan (CHN); Xingtai (CHN); Yongkang (CHN). 

\smallskip

\noindent 25 sample FUAs in \textbf{Europe}: Amsterdam (NLD); Athens (GRC); Barcelona (ESP); Belgrade (SRB); Cardiff (GBR); Chelyabinsk (RUS); Cologne (DEU); Dortmund (DEU); Glasgow (GBR); Helsinki (FIN); Katowice (POL); Krakow (POL); Kyiv (UKR); Leeds (GBR); Lisbon (PRT); Liège (BEL); London (GBR); Moscow (RUS); Nuremberg (DEU); Saratov (RUS); Seville (ESP); Sofia (BGR); Vienna (AUT); Voronezh (RUS); Warsaw (POL). 

\smallskip

\noindent 25 sample FUAs in \textbf{North America}: Cincinnati (USA); Dallas (USA); Guatemala City (GTM); Kansas City (USA); Las Vegas (USA); Managua (NIC); Montreal (CAN); Orlando (USA); Ottawa (CAN); Philadelphia (USA); Pittsburgh (USA); Raleigh (USA); Richmond (USA); Río Piedras [San Juan] (PRI); Sacramento (USA); Salt Lake City (USA); San Jose (USA); San Salvador (SLV); St. Louis (USA); Tegucigalpa (HND); Tijuana (USA); Toluca (MEX); Torreon (MEX); Vancouver (CAN); Washington D.C. (USA).

\smallskip

\noindent 6 sample FUAs in \textbf{Oceania}: Adelaide (AUS); Auckland (NZL); Brisbane (AUS); Melbourne (AUS); Perth (AUS); Sydney (AUS).

\smallskip
\noindent 25 sample FUAs in \textbf{South America}: Belo Horizonte (BRA); Belém (BRA); Bucaramanga (COL); Buenos Aires (ARG); Cali (COL); Cochabamba (BOL); Curitiba (BRA); Lima (PER); Maceió (BRA); Manaus (BRA); Maracaibo (VEN); Maracay (VEN); Medellín (COL); Mendoza (ARG); Natal (BRA); Porto Alegre (BRA); Quito (ECU); Recife (BRA); Rosario (ARG); Salvador (BRA); Santiago (CHL); Santos (BRA); São Luís (BRA); São Paulo (BRA); Valencia (VEN).

\subsection*{OSM query for street network data}
The custom filter used for the OSM query for street network data is as follows:
\begin{verbatim}
custom_filter = '["highway"~"living_street|motorway|motorway_link|
pedestrian|primary|primary_link|residential|secondary|
secondary_link|service|tertiary|tertiary_link|trunk|trunk_link
|unclassified"]' 
\end{verbatim}
Note that all streets with \texttt{highway=service} are filtered out at a later step, and that the network that is polygonized does not contain any service roads.

\subsection*{Technical specifications for peak/valley finding}
We fit a Gaussian kernel density estimation (KDE), using the parametric Silverman method for bandwidth selection. Then, we look for local maxima (peaks) and local minima (valleys) in the KDE with the following parameters: no minimum or maximum height for local minima; minimum height for local maxima set to $0.0008$ to exclude noise; no maximum height for local maxima; and a minimum peak prominence of $0.00075$. 

\subsection*{Pearson correlation coefficients between tested shape metrics}
\begin{table}
    \centering
    \begin{tabular}{p{0.25\textwidth}p{0.15\textwidth}p{0.15\textwidth}p{0.10\textwidth}p{0.10\textwidth}p{0.10\textwidth}}
    \toprule
     & circular \mbox{compactness} & isoperimetric quotient & isoareal \mbox{quotient} & radii ratio & diameter ratio \\
    \midrule
    circular compactness & 1.000 & 0.938 & 0.903 & 0.979 & 0.874 \\
    isoperimetric quotient & 0.938 & 1.000 & 0.987 & 0.958 & 0.765 \\
    isoareal quotient & 0.903 & 0.987 & 1.000 & 0.952 & 0.735 \\
    radii ratio & 0.979 & 0.958 & 0.952 & 1.000 & 0.858 \\
    diameter ratio & 0.874 & 0.765 & 0.735 & 0.858 & 1.000 \\
    \bottomrule
    \end{tabular}
    \caption{Pearson correlation coefficients between tested shape metrics.}
    \label{tab:pearson}
\end{table}

\subsection*{Spearman  correlation coefficients between tested shape metrics}
\begin{table}
    \centering
    \begin{tabular}{p{0.25\textwidth}p{0.15\textwidth}p{0.15\textwidth}p{0.10\textwidth}p{0.10\textwidth}p{0.10\textwidth}}
    \toprule
     & circular \mbox{compactness} & isoperimetric quotient & isoareal \mbox{quotient} & radii ratio & diameter ratio \\
    \midrule
    circular compactness & 1.000 & 0.952 & 0.952 & 1.000 & 0.872 \\
    isoperimetric quotient & 0.952 & 1.000 & 1.000 & 0.952 & 0.785 \\
    isoareal quotient & 0.952 & 1.000 & 1.000 & 0.952 & 0.785 \\
    radii ratio & 1.000 & 0.952 & 0.952 & 1.000 & 0.872 \\
    diameter ratio & 0.872 & 0.785 & 0.785 & 0.872 & 1.000 \\
    \bottomrule
    \end{tabular}
    \caption{Spearman  correlation coefficients between tested shape metrics.}
    \label{tab:spearman}
\end{table}

\end{document}